\documentstyle[11pt,aaspp4,psfig]{article} %preprint format

\def\deg{\ifmmode^\circ\else$^\circ$\fi}

\def\kpc{\ifmmode h^{-1}{\rm kpc}\else$h^{-1}{\rm kpc}$\fi}
\def\kms{\ifmmode {\rm km~s}^{-1}\else${\rm km~s}^{-1}$\fi}
\def\bii{\ifmmode b^II\else b$^{II}$\fi}
\def\lii{\ifmmode l^II\else l$^{II}$\fi}
\def\feh{\ifmmode {\rm [Fe/H]}\else [Fe/H]\fi}
\def\mvd{M_V({\rm D})}
\def\mvsg{M_V({\rm SG})}
\def\mvto{M_V({\rm TO})}

\lefthead{Beers et al.}
\righthead{Metal-Poor Stars in the Galaxy}

\begin{document}

\title{Kinematics of Metal-Poor Stars in the Galaxy. II. Proper Motions 
for a Large Non-Kinematically Selected Sample}

\author{Timothy C. Beers}
\affil{Department of Physics \& Astronomy, Michigan State University, E.
Lansing, MI 48824\\email:  beers@pa.msu.edu}
\author{Masashi Chiba}
\affil{National Astronomical Observatory, Mitaka, Tokyo 181-8588, Japan\\
email: chibams@gala.mtk.nao.ac.jp}
\author{Yuzuru Yoshii\altaffilmark{1}}
\affil{Institute of Astronomy, Faculty of Science, University of Tokyo,
Mitaka, Tokyo 181-8588, Japan
\\email: yoshii@ioa.s.u-tokyo.ac.jp}
\author{Imants Platais}
\affil{Department of Astronomy, Yale University, P.O. Box 208101, New Haven, CT
06520-8101\\email: imants@astro.yale.edu} 
\author{Robert B. Hanson}
\affil{ University of California Observatories / Lick Observatory, University
of California, Santa Cruz, CA  95064
\\email: hanson@ucolick.org}
\author{Burkhard Fuchs}
\affil{Astronomisches Rechen-Institut Heidelberg, M\"onchofstr. 12-14,
69120 Heidelberg, Germany\\email: fuchs@ari.uni-heidelberg.de}
\author{Silvia Rossi}
\affil{Instituto Astron\^omico e Geof\'isico, Universidade de S\~ao Paulo,
Brazil
\\email: rossi@orion.iagusp.usp.br}

\altaffiltext{1}{Also {\it RESCEU:} Research Center for the Early Universe,
Faculty of Science, University of Tokyo, Bunkyo-ku, Tokyo 113-0033, Japan}

\begin{abstract}

We present a revised catalog of 2106 Galactic stars, selected without
kinematic bias, and with available radial velocities, distance estimates, and
metal abundances in the range $0.0 \le \feh \le -4.0$.  This update of the
Beers \& Sommer-Larsen (1995) catalog includes newly-derived homogeneous
photometric distance estimates, revised radial velocities for a number of stars
with recently obtained high-resolution spectra, and refined metallicities for
stars originally identified in the HK objective-prism survey (which account for
nearly half of the catalog) based on a recent re-calibration.  A subset of 1258
stars in this catalog have available proper motions, based on measurements
obtained with the {\it Hipparcos} astrometry satellite, or taken from the
updated Astrographic Catalogue (AC 2000; second epoch positions from either the
Hubble Space Telescope Guide Star Catalog or the {\it Tycho} Catalogue), the
Yale/San Juan Southern Proper Motion (SPM) Catalog 2.0, and the Lick Northern
Proper Motion (NPM1) Catalog.  Our present catalog includes 388 RR Lyrae
variables (182 of which are newly added), 38 variables of other types, and 1680
non-variables, with distances in the range 0.1 to 40 kpc.

\end{abstract}


\section{Introduction}

Studies of the kinematics of various stellar populations in the Galaxy, in
particular the thick disk and the nascent halo, have long been limited by the
availability of large samples of stars with measurements of velocities,
distances, and metallicities.  Such a database is required in order to
constrain plausible scenarios for the formation and evolution of the Milky Way
and other large spiral galaxies like it.  Current issues which might be
addressed with such data include: (a) the rotational character of the thick
disk and halo (see, e.g., Beers \& Sommer-Larsen 1995, hereafter BSL, and
references therein), (b) the existence and the observed lower limit on the
metal abundance of stars in the so-called metal-weak thick disk (MWTD)
(Morrison, Flynn, \& Freeman 1990, hereafter MFF; BSL; Carney et al. 1996;
Chiba \& Yoshii 1998; Martin \& Morrison 1998) (c) a dual-component (flattened
plus spherical) halo population in the Galaxy (Hartwick 1987; Sommer-Larsen \&
Zhen 1990; Norris 1994; Kinman, Suntzeff, \& Kraft 1994; Sommer-Larsen et al.
1997), (d) quantitative estimates of the local density of thick disk and halo
stars (Yoshii 1982; Preston, Shectman, \& Beers 1991; Morrison 1993), (e) tests
for the existence of a putative ``counter-rotating'' halo component beyond 4--5
kpc from the plane (Majewski 1992; Wilhelm 1995; Carney et al.  1996; Wilhelm
et al.  1996; Zinn 1996; Carney 1999), (f) measures of the local halo velocity
ellipsoid for comparison with the derived ellipsoid for more distant halo stars
(Sommer-Larsen et al. 1997), and (g) derivation of a reliable RR Lyrae absolute
magnitude estimate based on statistical parallax analyses (Layden et al. 1996;
Fernley et al. 1998; Popowski \& Gould 1998).  As more extensive searches are
carried out for evidence of past (and present) mergers of smaller galaxies with
the Milky Way (e.g., Preston, Beers, \& Shectman 1994; Harding et al. 1998;
Helmi \& White 1999; Majewski 1999), it is of equal importance to obtain
secure knowledge of the ``global'' Galactic kinematic properties, so that
deviations from the expected behavior can be reliably assessed.

To confidently address the above issues (and many others) stars chosen as
kinematic tracers should be identified in a manner which does {\it not} depend
on a kinematic selection criterion itself.  Although it may be possible to
statistically correct for an input selection bias of this nature (Bahcall \&
Casertano 1986; Ryan \& Norris 1991a; Carney et al. 1994; Carney 1999), one is
left with lingering doubt concerning the derived kinematic parameters based on
post-facto modifications of the results.  It is similarly important that the
tracer stars cover a wide range of metallicities and distances (over both
northern and southern Galactic hemispheres), so that correlations of kinematics
as a function of these parameters can be investigated.

Although hints of the impact of a kinematic selection criterion are
evident in the work of Yoshii \& Saio (1979), the first {\it
large} database suitable for exploration of many of these issues was that of
Norris (1986), which included some 400 spectroscopically and/or photometrically
selected stars with abundances $\feh \le -0.6$, and with available radial
velocities and distance estimates.  In Paper I of this series BSL extended the
Norris catalog by inclusion of some 900 stars identified from the HK object
objective-prism survey of Beers \& colleagues (Beers, Preston, \& Shectman
1985; Beers, Preston, \& Shectman 1992a; Beers et al. 1992b), as well as some
600 additional stars from other smaller samples which appeared in the
literature subsequent to the publication of the Norris catalog, obtaining a
total sample of 1936 stars.

In this paper we present a revision of the BSL catalog of Paper I, based on
additional observational information which has recently become available.  In
addition to revisions of metallicities for the published HK
survey stars, new photometric distance estimates have been made for the entire
BSL catalog based on an internally self-consistent methodology, and in some
cases, new photometry.  Radial velocities have been updated based on
recently-obtained high-resolution data for a number of stars.  Much more
accurate positional information for the stars in our revised catalog has been
obtained by comparison with astrometric positions available for many of the
brighter stars in our sample, plus improved information from automated scans of
wide-field photographic plates, such as compiled in USNO-A V2.0 (Monet et al.
1998), NPM1 (Klemola, Hanson, \& Jones 1993), and SPM 2.0 (Platais et al.
1998).  We have also added 182 RR Lyrae variables from the recent work of
Layden (1994), Layden et al. (1996), and Fernley et al. (1998), so that useful
comparisons of the kinematics of these stars with the non-variables (which
presumably sample the same Galactic phase space) can be carried out.

The {\it major} difference between the present catalog and the BSL catalog is
the addition of proper motions, from a variety of sources, for over half of the
stars in our catalog.  In \S 2 we discuss the assemblage of the present
catalog.  Revisions of radial velocities and abundances, and in particular,
distance estimates, are discussed in \S 3.  In \S 4 we present the new proper
motion information, and discuss the averaging we have carried out in order to
minimize statistical errors.  In \S 5 we discuss the observed characteristics
of the stars in the revised catalog, and derive estimates of space motions and
orbital parameters for the subset of stars with complete kinematical
information .  In the accompanying analysis paper (Chiba \& Beers 2000; Paper
III) we use this wealth of new information to consider many of the questions
put forth above.

\section{Additions and Subtractions from the Beers \& Sommer-Larsen
(1995) Catalog}

The BSL catalog of Paper I contained 1936 stars with available abundances,
distances, and radial velocities.  As it was our intent to use this catalog as
the starting point in searches for stars with measured proper motions, the
first requirement was to obtain updated positions for the stars of as high an
accuracy as possible.  A small number of fainter stars in the BSL
catalog with crude positions (and no finding charts) were eliminated entirely.
The remaining stars were then compared with the positions listed in the SIMBAD
database.  A first search was done on the star name, followed by a search on
the reported position in the BSL catalog, with the requirement that the
magnitude reported in SIMBAD be commensurate with that of the target star.
Many of the brighter stars from the BSL catalog now have astrometric positions
from {\it Hipparcos}, which are accurate at the milli-arcsecond (mas) level.
For the stars of intermediate and fainter magnitudes, searches were conducted
on the BSL positions by comparison with the USNO-A 2.0 catalog, again with the
requirement that the magnitudes be commensurate.  The typical accuracy of
stellar positions for USNO-A 2.0 is at the 0\farcs 25 (one-sigma) level
(Deutsch 1999).  Further searches were conducted within the catalogs we use
below to find proper motions -- NPM1, SPM 2.0, STARNET, and ACT -- which
resulted in positional accuracies on the order of 0\farcs 2, 30 mas, 0\farcs
3, and 25 mas, respectively.  For a small number of stars no improved
positional information could be found, hence we retained the positions in the
original BSL catalog.

The {\it Hipparcos}, {\it Tycho}, NPM1, and SPM 2.0 catalogs included a number
of ``targeted'' RR Lyrae stars which were not included in the BSL catalog,
some of which have radial velocity and abundance estimates available in the
literature.  We thus added a total of 182 RR Lyrae stars to our revised
catalog, including many with metallicities greater than the nominal cutoff of
the BSL catalog.  A small number of non-variables which were not included in
the BSL catalog, but which have available radial velocities and abundance
estimates, were also added.  As a result of the revised abundances, described
in more detail below, a small number of HK survey stars which were originally
assigned metallicities $\feh \le -0.6$ presently have values above this limit.
These stars are noted in the revised catalog.

Column (1) of Table 1A lists the star names.  We have endeavored to follow the
IAU-recommended nomenclature, but in a few cases had to abbreviate the star
name in order to save space.  In any case, the positional identification is
usually unambiguous.  All adopted positions were updated to the ICRS (2000.0)
system.  Columns (2) and (3) of Table 1A list the right ascension and
declination of the adopted positions, respectively.  The source of the position
is indicated in column (4).

\section{Revisions of Abundance Estimates, Velocities, and Distances}

\subsection{Revisions of Abundance Estimates}

Roughly 45 \% of the stars in the revised catalog were originally identified
during the course of the HK objective-prism survey.  In the BSL catalog the
abundance estimates for these stars were obtained from a calibration of the
strength of the CaII K line as a function of $(B-V)_o$ color described in Beers
et al.  (1990).  This calibration is now known to suffer from several
deficiencies, the most worrisome of which is the fact that assigned abundances
in the range $\feh > -1.5$ are less than optimal as a result of the
saturation of the CaII K line, in particular at cooler temperatures.  Stars in
the metallicity regime $-2.0 \le \feh \le -1.0$ represent the transition
between the halo and thick disk populations, and in particular, may be members
of the MWTD component of the Galaxy, hence it is important that the abundances
be correctly estimated.  At the low end of the metallicity scale ($\feh < -2.5
$), the lack of available calibrators in the Beers et al. (1990) treatment
resulted in stellar abundance estimates which were, on the whole, somewhat
lower than has proven to be justified on the basis of recent high-resolution
spectroscopy, in particular for the hotter stars near the main-sequence turnoff
of an old halo population.

The re-calibration of metallicity estimates based on medium resolution
spectroscopy described by Beers et al. (1999) avoids, to a great extent, the
above difficulties.  This re-calibration makes use of an additional metallicity
estimator based on the Auto-Correlation Function of metallic lines in
a stellar spectrum (originally described by Ratnatunga \& Freeman 1989), which
provides a superior estimate of abundance to CaII K in the regime $\feh >
-1.0$.  The combination of the CaII K and ACF approaches implemented in Beers
et al. (1999) results in metallicity estimates over the interval $-4.0 \le \feh
\le 0.0$ with typical errors on the order of 0.1-0.2 dex, and no significant
systematic offsets.

Figure 1 is a comparison of the abundance determinations of the HK survey stars
in the revised catalog based on the new and old calibration.  Note that the
broad sloping region about the one-to-one line, which is dominated by stars
with $0.3 \le (B-V)_o \le 0.5$, indicates that the revised abundances for the
hotter HK survey stars are generally higher at the low metallicity end of the
scale, and lower at the high metallicity end of the scale, as compared to the
old calibration.  The abundance estimates for the cooler stars with $0.5 <
(B-V)_o \le 1.2$ have changed relatively less.

As part of the re-calibration effort, Table 7 of Beers et al. (1999) lists
averaged high-resolution abundance determinations from the literature for 551
stars.  For stars which also appear in the revised catalog, we adopted these
averaged abundance determinations.  

The catalog of Cayrel de Strobel et al. (1997) provides detailed information on
abundances for stars with determinations based on high-resolution spectroscopy.
Suitably averaged abundances for stars which did not appear in Beers et al.
(1999) were adopted for the revised catalog listing.  A number of stars from
MFF have been shown by Ryan \& Lambert (1995) to suffer from mis-estimated
abundances (generally too low) arising from a faulty calibration of DDO
photometry.  For MFF stars with high-resolution abundances obtained by Ryan \&
Lambert, we simply adopt their values.  For other MFF stars which are included
in the revised catalog, the abundances should be considered uncertain.

Metallicity estimates for the revised catalog are reported in column (5) of
Table 1A.  The source of the metallicity estimate is indicated in column (6).
Uncertain abundances are indicated with a ':' appended to the reported \feh\ .

\subsection{Revision of Radial Velocities}

Radial velocities of improved accuracy have been reported for a number of stars
in our revised catalog based on high-resolution spectroscopic follow-up, in
particular for the HK survey stars (McWilliam et al. 1995; Norris, Ryan, \&
Beers 1996).  The adopted radial velocities are reported in column (7) of Table
1A.  Errors on the radial velocity measurements are reported in column (8), and
the source of the radial velocity estimate is provided in column (9).  For the
HK survey stars with only medium-resolution spectroscopy available, we adopt a
conservative error estimate of 10 \kms\ (based on previous comparisons).  For
non-variable stars where the original authors did not report an error on the
velocity determination, we have adopted an error of 10 \kms .  For RR Lyrae
variables without reported radial velocity errors, we assign an error of 30
\kms , which accounts for the uncertainty in the systemic velocity arising from
limited phase coverage in spectra for a given variable (Smith 1995).

\subsection{Revision of Distance Estimates}

We have determined new photometric estimates of stellar distances for stars in
our catalog based on $M_V$ {\it vs.} $(B-V)_o$ relations for non-variable
stars, as described below, and an adopted $M_V$ {\it vs.} $\feh$ relation for
RR Lyrae variables.

\subsubsection{Apparent Magnitudes and Colors}

Our primary source of $V$-band apparent magnitudes and $B-V$ colors for
non-variable stars is the photometry obtained during the course of the HK
survey, if available (with typical accuracy in $V$ magnitudes and $B-V$ colors
in the range 0.005 to 0.01, respectively).  Additional photometric information
was taken from the SIMBAD database, the {\it Hipparcos} Catalogue, and the
calibrated photographic photometry of SPM 2.0, and GSC 1.2.  

There are several hundred HK survey stars in our catalog which presently lack
measured photometry.  For these stars we have estimated the $V$ magnitudes and
$B-V$ colors in the following manner.  An estimate of the $B$ magnitude is
obtained from GSC 1.2, where available.  An approximate de-reddened $(B-V)_o$
color is obtained from available medium resolution spectroscopy, based on the
calibration of the Balmer line index $HP2$ described by Beers et al. (1999).
This is then reddened using the listed estimated to reddening in the direction
toward the star, and the resulting $B-V$ is obtained.  This color is then used
to convert the GSC $B$ magnitude estimate to an approximate $V$ magnitude.
Given the multiple approximations in these procedures, these estimates should
be regarded with appropriate caution -- these stars are indicated in Table 1A
with parentheses around the $V$ and $B-V$ magnitude estimates.  In a few cases,
apparent magnitudes for HK survey stars that were not found in the GSC are
estimated from the intensity of the spectrum on the original prism plate, and
are indicated as such with brackets around the $V$ and $B-V$ magnitudes in the
table.

For a number of stars, $V$ magnitudes and $b-y$ colors are available from
Schuster et al. (1996), Schuster et al. (2000), or Anthony-Twarog et al.
(2000).  For these stars, we have listed an approximate $B-V$ color obtained by
adopting the transformation $B-V= 1.35\; (b-y)$, and indicate these cases with
parentheses about the approximate $B-V$ color in .  For RR Lyrae variables, the
intensity-averaged $V$ magnitudes, where available, are taken from either
Layden (1994), Layden et al. (1996), or Fernley et al.  (1998).  The adopted
photometric quantities are reported in columns (10) and (11) of Table 1A; the
source of the photometry is given in column (12); a reddening estimate (as
described below) is listed in column (13).

\subsubsection {Estimates of Reddening}

Our primary source of total reddening in the directions of the sample stars is
the new map of Schlegel, Finkbeiner, \& Davis (1998, hereafter SFD) based on
{\it COBE} and {\it IRAS} measurements of dust emission in the Galaxy. The SFD
map is superior to the widely used map of Burstein \& Heiles (1982, hereafter
BH) in its spatial resolution and its zero point (see also Gould \& Popowski
1998 for arguments in favor of the SFD map).  Note, however, that Arce \&
Goodman (1999) caution that the SFD map overestimates the reddening values when
the color excess $E_{B-V}$ is more than about 0.15 mags, where SFD's
calibration from dust column density to reddening may no longer be accurate.
For the small number of stars where this large reddening applies, we have
adopted estimates from the BH map, where available.  The few program stars in
such highly reddened directions are generally located at small distances from
the Sun, so that the effects of correction for reddening on distance estimates
are small.

We adopt a $V$-band absorption $A_V = 3.1 E_{B-V}$ and assume that the dust
layer has a scale height $h = 125$ pc. The reddening to a given star at
distance $D$ is reduced compared to the total reddening by a factor $\exp[- |D
\sin b|/h ]$, where $b$ is the galactic latitude.  An iteration procedure is
employed to obtain consistent values of $E_{B-V}$ and $D$.  The final adopted
reddening is listed in column (13) of Table 1A.

\subsubsection{Classification of Stellar Types}

Column (14) of Table 1A lists the classification of each stellar type.  We
follow the coding of BSL -- D: main-sequence dwarf star; A: main-sequence
A-type star; TO: main-sequence turnoff star; SG: subgiant star; G: giant star;
AGB: asymptotic giant branch star; FHB: field horizontal-branch star; V: 
variable stars other than the RR Lyrae type.  For the RR Lyraes, we use a code
'RRV' to distinguish them from other types of variable stars.  Most of the
classifications are taken from the BSL catalog, with the exception of a few
stars where we have adopted the classifications from Norris, Bessel, \& Pickles
(1995) (HD 13359: G; HD 17233: SG; HD 219221: D; HD 217808: G; CPD-62 394: G)
and Anthony-Twarog \& Twarog (1994) (HD 161770: TO; BD-01 1792: G; HD 17072:
AGB; HD 41667: G).  These revised classifications provide photometric distance
estimates which are more consistent with the {\it Hipparcos} parallaxes, as
described below.  Revised classifications are also provided for a number of the
HK survey stars, based on more recent observations.

There still remains some uncertainty in the classification of a number of stars
with relatively blue colors [$0.3 \le (B-V)_o \le 0.5$].  If the corresponding
stars have $(U-B)_o$ color data from the HK survey, the stars are classified as
FHB if $(U-B)_o > -0.1$, or TO if $(U-B)_o \le -0.1$ (Wilhelm, Beers, \& Gray
1999).  This crude discrimination scheme is adequate for stars with abundances
$\feh<-1.5$.  For more metal-rich stars, we also consider the alternative
classification as possible, and assign a classification of FHB/TO.

\subsubsection{Templates for Distance Estimation}

RR Lyrae distance scales have been reconsidered extensively subsequent to the
release of the {\it Hipparcos} Catalogue, based on a variety of methods
including main-sequence fitting, the Baade-Wesselink method, and statistical
parallax.  Combining all of the results, Chaboyer (1999) proposed the following
$M_V$ {\it vs.} $\feh$ relation:

\begin{equation}
M_V (RR) = 0.23 ( \feh + 1.6 ) + 0.56 \ ,
\end{equation}

\noindent where $M_V (RR)$ denotes the absolute magnitude of RR Lyrae stars.
We adopt this relation to estimate the distances of our RR Lyrae sample.

For non-variable stars, we first calibrate fiducial points in the $M_V$ vs
$(B-V)_o$ relation, using published data from various Galactic globular
clusters and the Hyades and Pleiades open clusters to cover large ranges of
colors and metal abundances.  The globular clusters we use include M92
($\feh=-2.24$), M15 ($-2.15$), M3 ($-1.66$), M13 ($-1.65$), NGC~6752 ($-1.54$),
NGC~362 ($-1.27$), and 47~Tuc ($-0.71$) (Sandage 1970; Buonanno, Corsi, \& Fusi
Pecci 1989; Durrell \& Harris 1993; Penny \& Dickens 1986; Harris 1982; Hesser
et al.  1987), where metal abundances and extinctions are taken from the
compilation of Chaboyer, Demarque, \& Sarajedini (1996).  The absolute
magnitude of the horizontal branch in each cluster is scaled according to Eqn.
(1).  For the Hyades cluster ($\feh=+0.12$), we adopt the work of Perryman et
al. (1998), which utilized the {\it Hipparcos} parallaxes to determine the
distances to the cluster stars. Using their Figure 21, we read off the fiducial
$M_V$ vs $(B-V)_o$ relation for the stars which are not classified as double or
multiple in the {\it Hipparcos} Catalogue.  For the Pleiades cluster
($\feh=+0.11$, essentially identical to the Hyades abundance), we adopt the
$M_V$ vs $(B-V)_o$ relation published by Mermilliod (1981).

Next, polynomial fits to the $M_V$ vs $(B-V)_o$ relation are made for each type
of star; the fitting formula and the coefficients of the fits are provided in
Table 2.  For G, SG, and AGB stars with abundances $\feh < -2.5$, we use the
tabulated relation for $\feh=-2.5$. For D stars we employ an
interpolation/extrapolation procedure among the tabulated relations.  Figure 2a
demonstrates the adopted $M_V$ vs $(B-V)_o$ relations of G, SG, and AGB stars
at metal abundances appropriate for M92$+$M15 ($\feh=-2.20$), M13 ($-1.65$) and
47~Tuc ($-0.71$), whereas Figure 2b shows those of D stars for M92$+$M15,
47~Tuc, and Hyades.  Unfortunately, there remains some additional uncertainty
for TO stars, due to their positions at the ``curved'' part of the adopted
$M_V$ vs $(B-V)_o$ relation. We employ the following approximate method.  For a
given $(B-V)_o$ color, $M_V$'s assuming both a D and SG classification are
first estimated. If $\mvd$ is brighter than $\mvsg$, $\mvd$ is adopted as
$\mvto$.  If this is not the case, the intermediate $M_V$ which locates between
$\mvd$ and $\mvsg$ is calibrated using the formula given in Table 2.  This
formula is designed to reproduce the turnoff point from neighboring points of
the isochrones used in Beers et al. (1999). 

In order to estimate distances for the hotter main-sequence gravity A-type
stars (i.e., those with colors $-0.10 \le (B-V)_o \le 0.34$) we apply a small
vertical shift to the $M_V$ vs $(B-V)_o$ relation for the Pleiades so that it
matches that of the Hyades at $(B-V)_o = 0.34$.  To account for the
expected decline in luminosity with decreasing abundance, the adjusted Pleiades
line is corrected by applying shifts of the same size as those derived for the
cooler stars on the main sequence of the metal-poor globulars as compared to
the Hyades.  We note that this is only one of several choices we could have
made, and it may not be the optimal one, but it should serve for the present
purpose.  

Taking all of the above information into account, distances are then estimated
by means of an iteration procedure which includes modification of the
interstellar reddening.  These distances, $D_{pho}$, are indicated in column
(2) of Table 1B.  In some cases there was no reliable photometry or
classification available, and, occasionally, our iteration scheme failed to
converge.  In these circumstances we made use of the distances provided in the
original BSL catalog (who, in such cases, adopted the distance estimates of the
authors who provided the abundance and radial velocity information), listed as
$D_{BSL}$ in column (5) of Table 1B.  For the stars with ambiguous luminosity
classifications, we list two alternative distances in column (2).

The small number of stars with uncertain type classifications (either due to
the FHB/TO ambiguity noted above, or for the TO stars that appear too blue for
their adopted type designation when compared with an old isochrone) are noted
as such in column 12 of Table 1B, and are ultimately excluded from further
analyses.

\subsubsection {{\it Comparison of Photometric and Astrometric Distance 
Scales}}

There are 508 stars in our present catalog which were observed with the
{\it Hipparcos} astrometric satellite (ESA 1997).  To assess the reliability 
of our photometric distances, $D_{pho}$, obtained above, we plot in Figure 3 
the {\it Hipparcos} trigonometric parallaxes, $\pi_{HIP}$, versus our 
photometric parallaxes, $\pi_{pho}=1/D_{pho}$, for 508 stars with available 
photometric distances.  Plotting parallaxes, rather than distances, in 
Figure 3 has several advantages.  All the trigonometric parallaxes, even small
or negative values, can be used in an unbiased way (see the discussion on 
pp.~27--28 of Arenou \& Luri 1999).  Systematic errors in the data can be 
detected; an incorrect photometric distance scale would be seen as a slope 
not equal to unity; a zero-point error in the trigonometric parallaxes as a 
non-zero intercept.  For clarity, we have not plotted error bars in Figure 3; 
the mean error of $\pi_{HIP}$ is $\sim 1.6$~mas; the mean error of $\pi_{pho}$
is $\leq 20\%$ of $\pi_{pho}$.

The general trend of the data in Figure 3 seems to fall slightly below the 
45-degree line, indicating that $\pi_{pho}$ may be $\sim 20\%$ too large for 
the nearer stars in the BSL sample.  This largely reflects the fact that 
these nearby stars have the largest errors in $\pi_{pho}$.  The bulk of 
the data, especially the more distant stars, with smaller $\pi_{pho}$ errors, 
falls closer to the 45-degree line.  Thus we conclude from Figure 3 that there 
is not any significant systematic error in our photometric distance scale.
It is also worth noting that Figure 3 shows very few outliers, {\it i.e.}
cases where $\pi_{pho}$ and $\pi_{HIP}$ badly disagree.  The large spread in
$\pi_{HIP}$ near $\pi_{pho} = 0$ is due to the large errors in $\pi_{HIP}$
(up to 7~mas) for these faint, distant stars.

For the 424 stars with non-negative {\it Hipparcos} parallaxes, Table 1B 
lists in column (3) the {\it Hipparcos} distance estimates 
$D_{HIP} = 1/\pi_{HIP}$  and in column (4) their relative precisions
$\sigma_{\pi_{HIP}}/\pi_{HIP}$.  Because the photometric distances for the
nearest stars in our catalog have lower precision than the trigonometric
parallaxes, we chose to adopt the {\it Hipparcos} distances for 44 stars with
$\sigma_{\pi_{HIP}}/\pi_{HIP} < 0.12$.  The final adopted distance estimates
for our program stars are listed as $D_{adopt}$ in column (6) of Table 1B.

\section{Proper Motions}

\subsection{Sources of Proper Motions}

\subsubsection{The {\it Hipparcos} Catalogue}

The {\it Hipparcos} Catalogue is the primary result of the first space mission
dedicated to astrometry (ESA 1997), and it provides high accuracy proper
motions for 118,218 stars with $V < 12.5$ mag covering the entire sky.  In the
present sample, 508 stars are included in this catalog with average accuracies
of 1.61 mas~yr$^{-1}$ in $\mu_{\alpha^\ast}$ ($=\mu_\alpha \cos\delta$) and
1.29 mas~yr$^{-1}$ in $\mu_\delta$.  Kinematical analyses for a subset of these
{\it Hipparcos} stars are reported in Chiba \& Yoshii (1998) and Chiba, Yoshii,
\& Beers (1999).

\subsubsection{The SPM Catalog 2.0}

The Yale/San Juan Southern Proper Motion (SPM) Catalog 2.0 provides positions,
absolute proper motions and photographic $B,V$ magnitudes for 321,608 objects
with $5 < V < 18.5$, mainly in the declination zone from $-37\deg$ to
$-27\deg$, except within the Galactic zone of avoidance. In this declination
zone an effort was made to measure {\it all} HK survey stars from a database
provided by T. Beers prior to undertaking measurements of the plates.  A total
of 505 of our program stars have proper motions extracted from the SPM 2.0
Catalog, using a 10\arcsec\ search radius centered on the catalogued positions,
and satisfying the requirement that the apparent magnitudes be within 1 mag in
$B$ or $V$ of those adopted in Table 1A.  The description of the catalog
properties can be find in Platais et al. (1998), however, this paper addresses
only the separate catalog at the South Galactic Pole.  For additional details
of the SPM Catalog 2.0, the reader is referred to the World Wide Web at URL
{\tt http://www.astro.yale.edu/astrom/}.  The average accuracy of SPM proper
motions for the subset of 508 stars is 2.9 mas yr$^{-1}$.

\subsubsection{The Lick NPM1 Catalog}

The Lick Northern Proper Motion (NPM) program (Hanson 1997) is a photographic
survey measuring absolute proper motions, on an inertial system defined by
50,000 faint galaxies, for over 300,000 stars with $8 < B < 18$, covering the
northern two-thirds of the sky ($\delta > -23\deg $).  Part I of the NPM
program (``NPM1''), outside the disk of the Milky Way ($|b| > 10$ deg), was
completed in 1993; the Lick NPM1 Catalog (Klemola, Hanson, and Jones 1993)
contains 148,940 stars, with proper motions accurate to 5 mas/yr, and positions
accurate to 0\farcs 2 or better.  Part II of the NPM program (``NPM2''), in
the Milky Way sky, will be completed in roughly three years.

We searched for program stars within a $\pm 10\arcsec \times 10\arcsec $ error box, centered
on the NPM1 catalog positions (updated to epoch 2000.0 using the NPM1 proper
motions).  This procedure found matches for 241 stars; 95 \% of these comprise
a tight core with RMS position difference 0\farcs 5 in each coordinate, and can be
considered unambiguous matches.  The remaining 5 \% can be considered highly
probable matches; agreement between the cataloged and NPM1 apparent magnitudes
(better than $\pm \sim 1$ mag) confirms these.

NPM1 proper motions for an additional 58 RR Lyrae stars, from the list of
Layden et al. (1996), were added to our catalogue, making a total of 299
stars with NPM1 proper motion data.

\subsubsection{The STARNET Catalog}

The proper motions of the STARNET (STN) catalog were determined using the
Astrographic Catalogue (AC) as the first epoch positional source, and the
astrometrically upgraded Guide Star Catalog (GSC 1.2) as the second epoch
source. The AC has been made available in machine readable form by Nesterov et
al. (1990). Its reduction to the FK5 system was performed by S. R\"oser at the
Astronomisches Rechen-Institut (ARI).  The astrometrical upgrading of the Guide
Star Catalog is described in detail by Morrison et al. (1996). GSC 1.2 is
presently available at the Astronomical Data Center.  The result of combining
both catalogs is the STARNET catalog of proper motions (R\"oser 1996). It
contains about 4.3 million stars with an average density of 100 stars per
square degree.  The median magnitudes of the stars are $B = 12.0$ mag in the
southern hemisphere and $V = 11.5$ in the northern hemisphere, respectively,
with the faintest stars reaching $B,V \sim 13.5$.  The present rms accuracy of
STARNET positions is 0\farcs 3.  The accuracy of the proper motions, which have
been determined simply from combining the estimated positional errors at both
epochs, are about 5 mas yr$^{-1}$. The catalog is reduced to the FK5 reference
system and thus contains some small systematic differences with respect to ICRS
(see Vol. 3, Chapter 19 of the Hipparcos and Tycho Catalogues, ESA 1997), in
addition to some residual offsets still present in the GSC 1.2 positions.
STARNET is as yet unpublished, but is accessible upon request at ARI.

We performed searches for our program stars in STARNET using windows of
40\arcsec\  diameter centered on the sample stars, and obtained 774
identifications in STARNET, with an average accuracy of the proper motions of
4.7 mas yr$^{-1}$ in each component, respectively.  Since the STARNET catalogue
may contain mis-identifications of stars from first epoch to second, proper
motions larger than 200 mas yr$^{-1}$ in one component may be suspect.  We
therefore inspected such proper motions individually.  As it turned out all but
one of these stars were listed in the Hipparcos Catalogue; the remaining star
could be found in the PPM catalog (R\"oser and Bastian 1991; Bastian et al.
1993).  The agreement between the STARNET motions and the other measurements is
usually quite good, with deviations of the order of the expected accuracy of
the proper motions.

\subsubsection{The ACT Reference Catalog}

This recently completed catalog by the USNO contains positions and proper
motions for 988,758 stars with $V < 11 \sim 11.5$ mag covering the entire sky
(Urban, Corbin, \& Wycoff 1998a).  Accurate proper motions were calculated by
combining the positions from the {\it Tycho} Catalogue (ESA 1997) with those
from new reductions of the Astrographic Catalogue, referred to as AC 2000
(Urban et al. 1998b).  We performed a search of our sample stars within a
$10''$ radius centered at the positions of the cataloged stars, and obtained
unambiguous matches for 525 stars.  Average accuracies in proper motions are
3.1 mas yr$^{-1}$ and 2.9 mas yr$^{-1}$ in $\mu_{\alpha^\ast}$ and
$\mu_\delta$, respectively.

\subsection {Comparison of Proper Motions for Stars with Multiple Measures}

To assess the quality of the adopted proper motions, Figure 4 shows
the difference between the {\it Hipparcos} Catalogue and other ground-based
measurements for each proper-motion component, for the stars in common between
the catalogs.  While the {\it Hipparcos} errors are smallest in general, the figures
show that there is an overall agreement with proper motion measurements from the
additional sources, without any clearly systematic differences as compared with
{\it Hipparcos}. It is worth noting that for several stars, the proper motion
accuracy in the SPM~2.0 Catalog is better than that in the {\it Hipparcos}
Catalogue.

For the stars which have been independently measured in two or more catalogs
it is possible, by combining all measurements, to reduce the statistical errors
as well as minimize any small remaining systematic errors in the individual
catalogs, as was argued in Martin \& Morrison (1998).  For these stars, we
estimate average proper motions, $<\mu>$, and their errors, $<\sigma_\mu>$,
weighted by the inverse variances, as

\begin{eqnarray}
<\mu> &=& (\sum_{i=1}^{n} \mu_i/\sigma_{\mu_i}^2)
      / (\sum_{i=1}^{n} 1/\sigma_{\mu_i}^2) \ , \\
<\sigma_\mu> &=& (\sum_{i=1}^{n} 1/\sigma_{\mu_i}^2)^{-1/2} \ ,
\end{eqnarray}

\noindent where $n$ denotes the number of measurements.  In this manner, the
proper motions for 691 stars in the present sample are improved.  Our adopted
proper motions  are listed in columns (7) and (8) of Table 1B; the associated
errors are listed in columns (9) and (10).  The sources of the proper motions
are listed in column (11).

We emphasize the large number of the stars having proper motion data in
the present catalog: $N = 1258$ (1214 with $\feh \le -0.6$). This is the {\it
largest} such dataset among any non-kinematically selected samples, and is
substantially larger than a number of recent samples selected on the basis of
their high proper motions alone (Ryan \& Norris 1991a: $N = 727$ with
$\feh\le-0.6$; Carney et al. 1994: $N = 689$ with $\feh\le-0.6$).  Furthermore,
the quality of the proper motion data has been improved in a meaningful manner,
compared with that of earlier studies, due to the continuous improvement of
astrometric observations from the ground as well as from space.  As a
consequence the current database allows us to elucidate a great deal of
information concerning the space motions of the metal-poor populations of the
Milky Way with a lowest-to-date observational error.

\section{The Revised Catalog and its Characteristics}

\subsection{The Catalog}

Tables 1A and 1B present the final revised catalog for 2106 stars. With a
constraint of $\feh \le -0.6$, the present catalog contains 2041 stars,
including 356 RR Lyrae variables (168 of which are newly added as compared to
the BSL catalog of Paper I), 38 variables of other types, and 1647
non-variables.  Proper motions are available for 1214 stars with $\feh
\le -0.6$.

\subsection{Characteristics of the Revised Catalog}

Figure 5a illustrates the distribution of apparent magnitudes for stars in the
present catalog.  The relatively small number of stars with magnitudes $V >
15$, which, for the giants of metal-poor populations, are located at distances
in excess of 10 kpc from the Sun, clearly points to the need for future work.
The distribution of $(B-V)_o$ colors for our catalog stars is shown in Figure
5b.  The dominance of D and MS stars is clear, pointing out the need for
additional studies of redder G  and AGB stars of the thick disk and halo, as
well as the need for the inclusion of blue FHB stars.

Figure 5c shows the distribution of adopted distances for stars in the present
catalog. Stars populating the inner few kpc from the Sun are well
represented, with a rather sharp decline beyond 3 kpc from the Sun.  The
distribution of revised stellar abundances shown in Figure 5d underscores the
fact that we have included a large number of stars with abundances {\it below}
the peak metallicity of the halo metallicity distribution function ([Fe/H]
$\simeq -1.6$, Laird et al. 1988; Ryan \& Norris 1991b).  Future work which
supplements this sample with additional stars in the intermediate abundance
range $-1.5 \le \feh \le -0.6$ is clearly needed.

The dashed histograms in Figures 5a-d indicate the distribution of the above
observables for the subset of stars with available proper motions.  As is
evident from inspection of Figure 5a, the great majority of the brighter
stars have measured proper motions, but the completeness falls from about 50\%
to roughly 30\% for the stars in the magnitude range $12 \le V \le 16$.  Figure
5c shows that the fraction of stars with available proper motions is
roughly a constant $\sim$ 60\% out to distances on the order of 4--5 kpc, with
no strong dependence on distance.

In Figures 6a and 6b we show the distribution of adopted radial velocities
(for all catalog stars with [Fe/H] $\le -0.6$) and measured proper motions
(where available) as a function of the stellar abundances.  The largest radial
velocities and proper motions are observed near the center of the abundance
distribution, but this is presumably a result of the (presently unavoidable)
fact that the relatively rare large observed motions will occur with greatest
frequency among the stars which are most commonly represented in the catalog.

The referee has raised the question of whether we have ``imported'' a kinematic
bias into the sample of stars with measured proper motions by the addition of
information from the sources we have drawn on.  This is demonstrably {\it not}
the case.  None of the five catalogs we have used in our
assemblage of proper motion information have lower limits on their proper
motions (as do, for example, many older catalogs such as those from Giclas and
Luyten).  The stars in each of these catalogs are chosen in advance of
measuring their proper motions, hence a given star is not rejected if its
measured proper motion turns out to be small, even if it is consistent with
zero.  Furthermore, since the proper motions we have included are measured in
two {\it independent} coordinates, each of which can be individually negative,
zero, or positive, errors in the measured proper motions do not result in a net
bias away from zero.  This bias does exist in many of the older catalogs (where
selection was based on the {\it always positive} length of a tangential motion
vector), but it does not apply to the catalogs we have used for our proper
motions. 

Since the above biases could lead, in effect, to a distance-related bias in the
resulting kinematics, we have performed a simple diagnostic to demonstrate that
this is not the case for our sample of stars (other tests are described in
Paper III).  Figure 6c shows a plot of the estimated tangential velocity of our
sample stars as a function of distance from the Sun.  Superposed on this figure
are lines of constant proper motion, corresponding to 10 mas/yr and 1 mas/yr,
respectively.  There does not appear to be any obvious changes in the
distribution of predicted tangential velocity ($V_t = 4.74\; \mu \; D$, where
$D$ is the distance in kpc, and $\mu$ is in mas/yr) that correlate with
increasing distance from the Sun, as might be expected if there were any
``hidden'' kinematic bias in the sample of stars with measured proper motions.

The spatial distribution of the sample in $(R,Z)$ plane is shown in Figure 7,
where filled circles and crosses indicate the stars with and without available
proper motions, respectively.  Below the Galactic plane there are many remote
stars with available proper motions, mainly provided by SPM 2.0.

\subsection{Full Space Motions and Orbital Parameters}

We now obtain estimates of the full space motions for the subsample of stars
for which proper motions are available, and derive parameters of their orbital
motions in a given gravitational potential.  The results are summarized in
Table 3.  Note that two of the stars in this subsample were judged not to have
sufficiently accurate distance estimates to derive space motions, V* KR Vir
and BPS CS 29512-0032, so these stars are not included in Table 3.  Column (1)
lists the star name.  Column (2) recalls the adopted metallicity from Table 1A.
Columns (3) and (4) list the positions of the stars in the meridional plane
$(R,Z)$, adopting $R_\odot = 8.5$ kpc as the Galactocentric distance for the
Sun.  Columns (5)-(7) list the three-dimensional velocities $U$, $V$, and $W$,
directed to the Galactic anticenter, the rotational direction, and the north
pole, respectively.  These velocity components are corrected for the solar
motion $(U_\odot,V_\odot,W_\odot)=(-9,12,7)$ km s$^{-1}$ with respect to the
local standard of rest (LSR) (Mihalas \& Binney 1981).  Figure 8 is a plot of
the $U,V,W$ velocity components for the sample of stars with complete velocity
information and $\feh \le -0.6$, as a function of \feh .

Columns (8) and (9) of Table (3) list the velocity components $(V_R,V_\phi)$ in
the cylindrical rest frame $(R,\phi)$, respectively, on the assumption that the
rotational speed of the LSR around the Galactic center is $V_{LSR} = 220$ km
s$^{-1}$. Note that for the stars for which accurate distances are not
available, columns (2)-(9) contain no data.

To estimate the orbital parameters for these stars, we adopt the analytic
St\"ackel-type mass model developed by Sommer-Larsen \& Zhen (1990), which
consists of a flattened, oblate disk and a nearly spherical massive halo. This
model reproduces a flat rotation curve beyond $R=4$ kpc and the local mass
density at $R_\odot$ in a reasonable manner. Its analytic form has the great
advantage of maintaining clarity in further analyses (see, e.g., Chiba \&
Yoshii 1998).  Columns (10) and (11) list the estimated apogalactic distances,
$R_{ap}$, and the estimated perigalactic distances, $R_{pr}$ along the Galactic
plane, respectively.  Column (12) lists the maximum distance above (or below)
the plane, $Z_{max}$, explored by each star in the course of its orbital
motion.  In column (13) we list the characteristic eccentricities of the
orbits, defined as $e = (r_{ap}-r_{pr})/(r_{ap}+r_{pr})$, where $r_{ap}$ and
$r_{pr}$ stand for the apogalactic and perigalactic distances from the Galactic
center, respectively. We note that the current mass model of the Galaxy fails
to gravitationally bind the 16 stars which lack the data of orbital parameters
in columns (10)-(13). These stars probably correspond to those in the ``error
tail'' of overestimated photometric distances, thereby having overestimated
transverse velocities.

In the accompanying analysis paper (Chiba \& Beers 2000; Paper III),
we make use of the kinematic information in Table 3, and the radial velocity
and distance information for the stars presently without available proper
motions, to investigate a number of questions concerning the nature of
the kinematics of the Galaxy.

\subsection{Future Samples}

Exploration of the kinematic properties of the Galaxy is very much a work in
progress.  New and extensive samples of several thousand additional
non-kinematically selected metal-deficient stars with available velocity,
distance, and abundance measurements are expected to become available within
the next year (Beers et al. 2000; Cayrel et al. 2000; Norris et al. 2000;
Rebolo et al. 2000). Roughly 30\% of these stars already have available proper
motions; completing the search for HK survey stars with proper motions awaits
the extension of the SPM catalog to other areas of the southern Galactic
hemisphere.

Samples of intermediate distance ($2 \le d \le 15$ kpc) FHB and other A-type
stars (e.g., Wilhelm et al.  1999), supplemented with observations of more
distant metal-poor dwarfs, giants, and FHB and A-type stars from the
Hamburg/ESO survey ($10 \le d \le 25$ kpc; Christlieb 1999), and from other
sources (e.g., the Sloan Digital Sky Survey, Pier 1999), will provide useful
extensions of our catalog.

A project to refine systemic radial velocity measurements for the RR Lyraes
with poorly determined values is already underway.  Additional $UBV$ and/or
Str\"omgren photometry is required in order to resolve ambiguities in stellar
classifications which remain in the catalog, and to obtain more precise
estimates of distances, especially for the stars with available proper motions.

\acknowledgments

We are grateful to the referee, Bruce Carney, for a careful reading of this
manuscript, and for a number of thoughtful comments.  

TCB acknowledges support for this work from grant AST 95-29454 from the
National Science Foundation.  MC and YY acknowledge partial support from
Grants-in-Aid for Scientific Research (09640328) and COE Research (07CE2002) of
the Ministry of Education, Science, Sports and Culture of Japan.  MC thanks
Hideyuki Saio for his help in distance estimates for the current catalog stars.
The SPM program is supported by grants from the NSF to Yale University and the
Yale Southern Observatory, Inc.  IP thanks T. Girard, V.  Kozhurina-Platais,
and W. van Altena for their expertise and contribution to the SPM program.  RBH
thanks the National Science Foundation for its long-term support of the Lick
Northern Proper Motion program.  Current work on the NPM program is supported
by NSF grant AST 95-30632.  RBH thanks A. Klemola for his help in providing
identifications for the Lick NPM1 Catalog stars.  BF thanks to S. R\"oser
and S. Frink for help with the STARNET catalog.  SR acknowledges partial
support for this work from grant 200068/95-4 CNPq, Brazil, and from the
Brazilian Agency FAPESP.

This work made use of the SIMBAD database, operated at CDS, Strasbourg, France.

\clearpage
%%%%%%%%%%%%%%%%%%%%%%%%%%%%%%%%%%%%

\centerline{\psfig{file=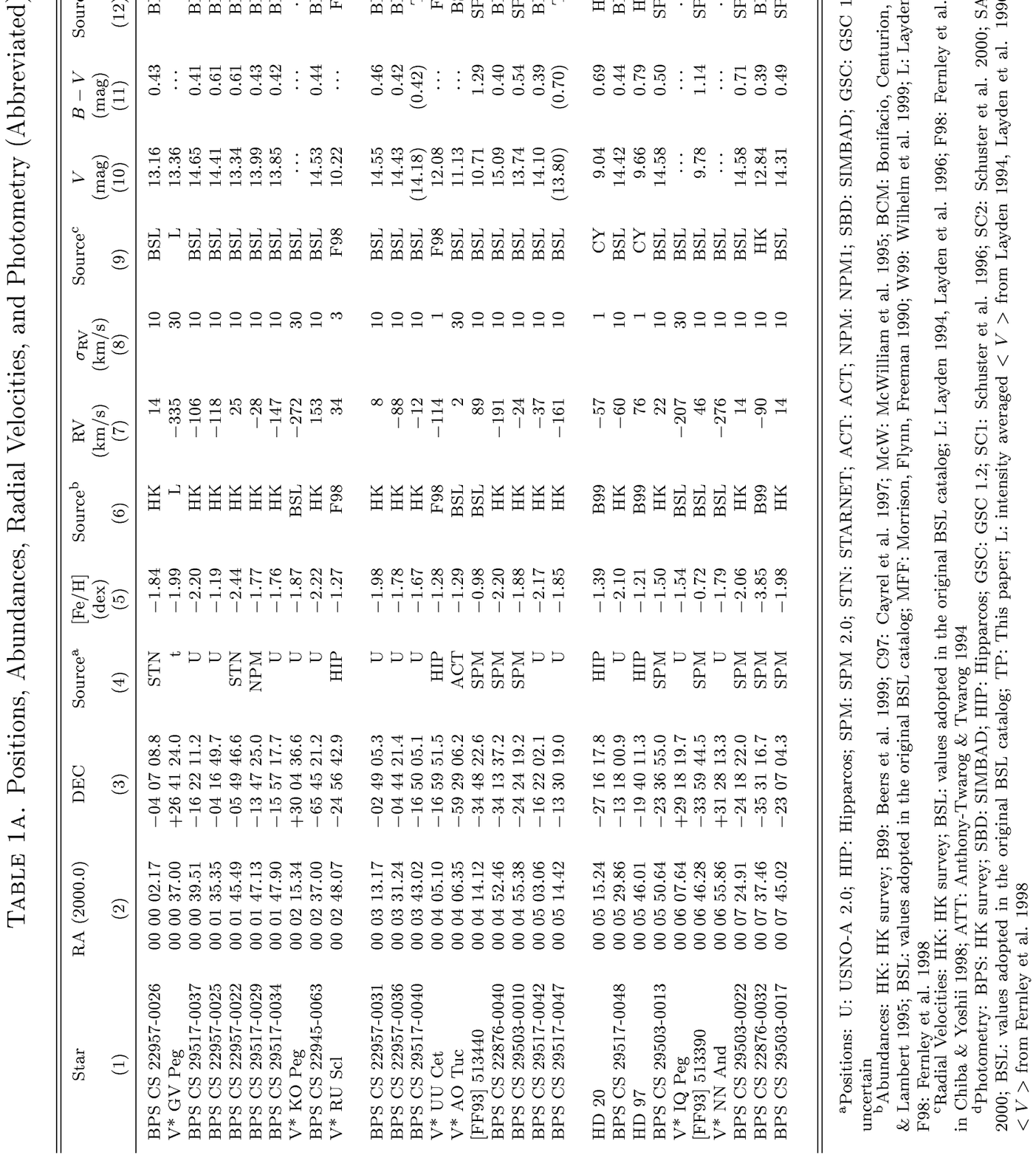,height=8.0in,angle=90}}
\centerline{\psfig{file=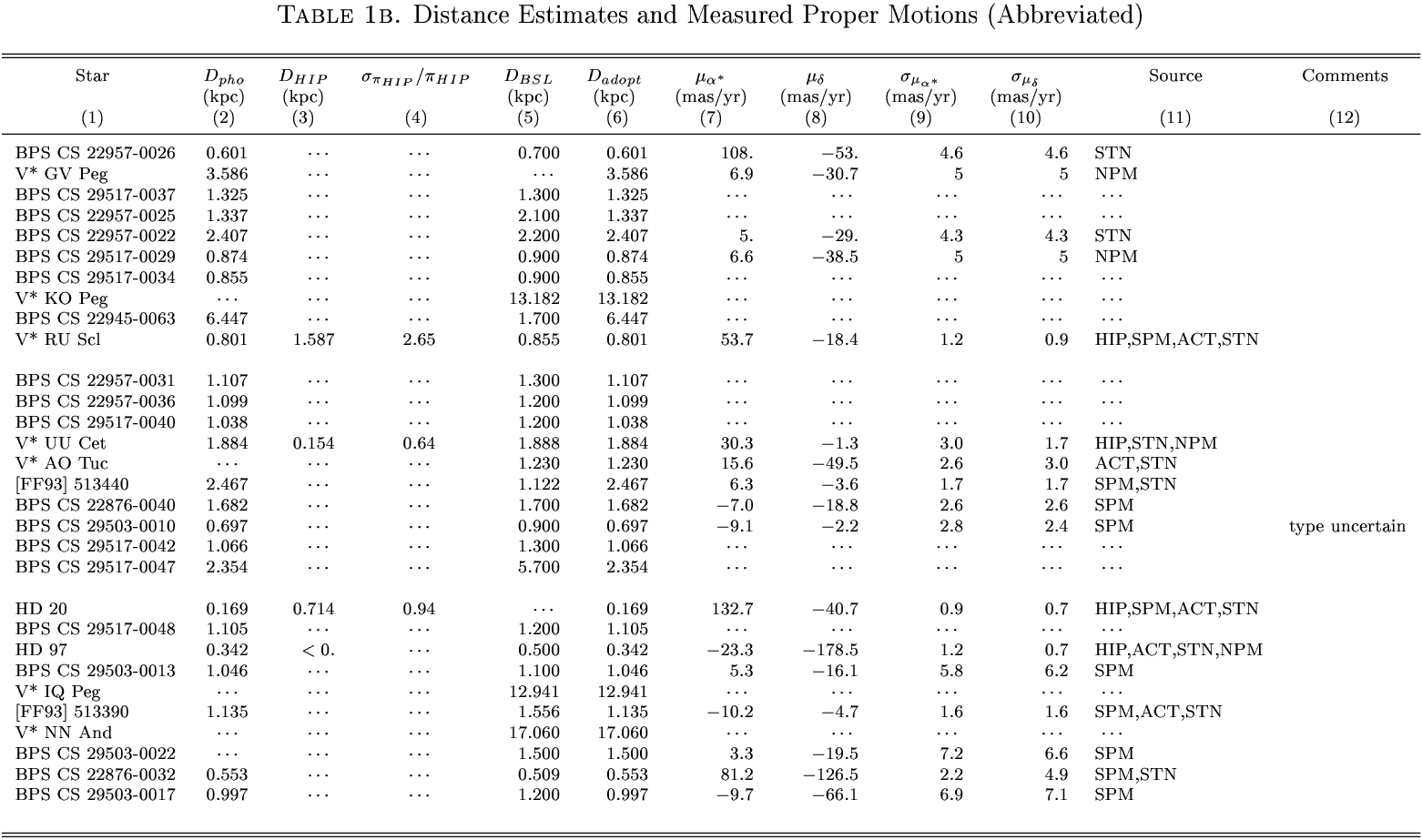,height=8.0in,angle=90}}
\centerline{\psfig{file=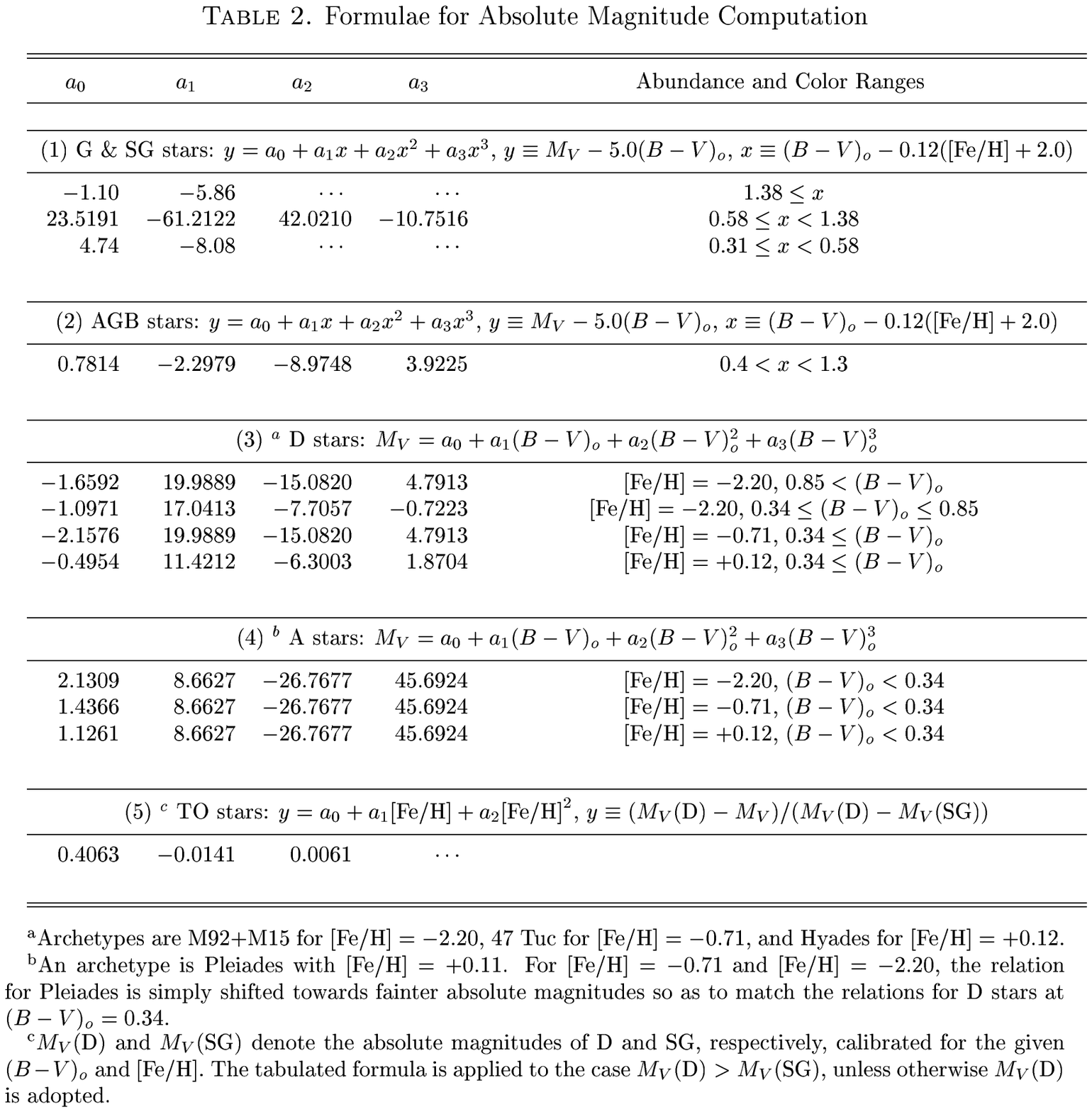}}
\centerline{\psfig{file=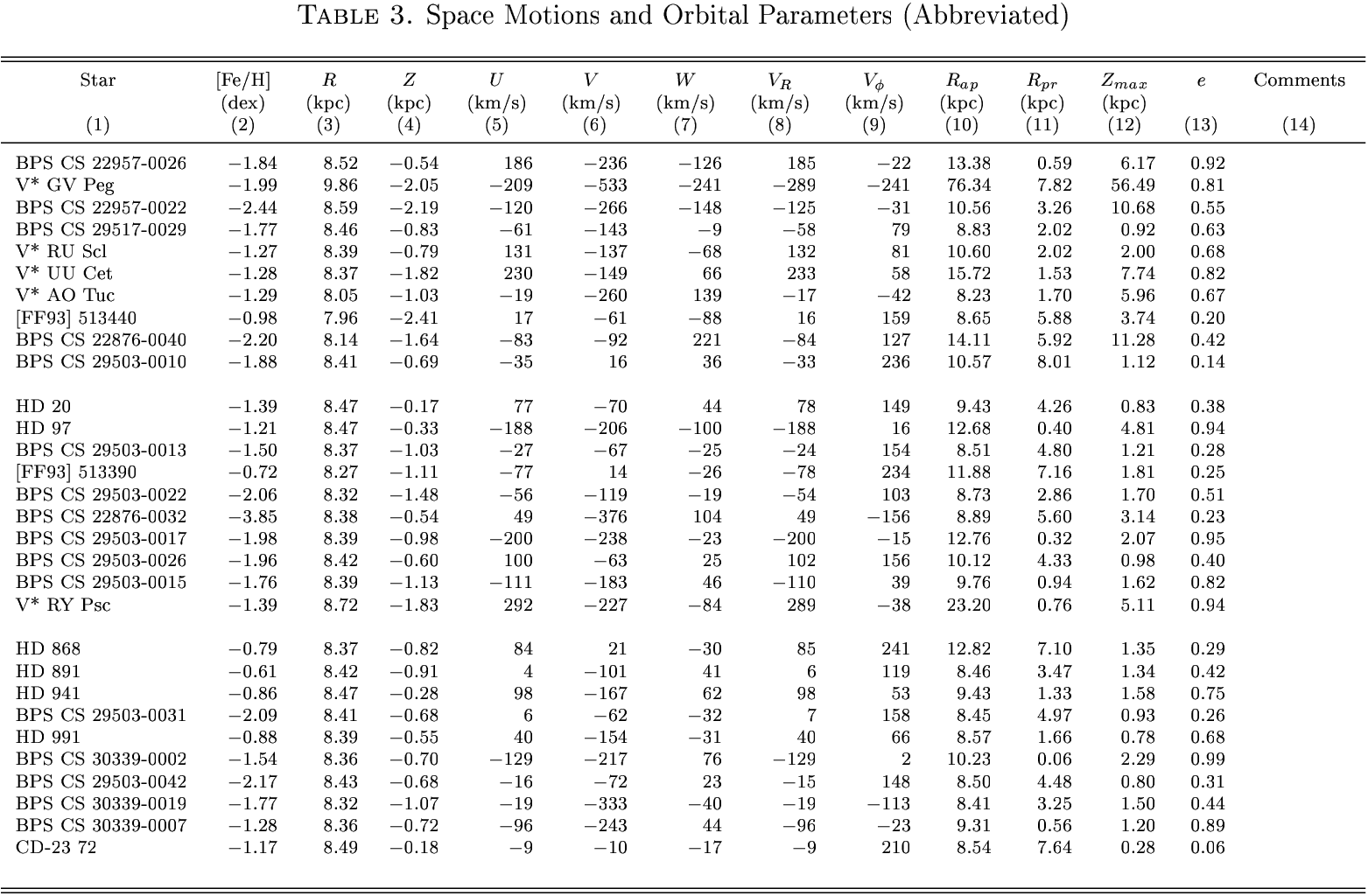,height=8.0in,angle=90}}

\clearpage
%%%%%%%%%%%%%%%%%%%%%%%%%%%%%%%%%%%%
\figurenum{4}
\figcaption[]{
Distribution of the differences $\Delta\mu$ between
{\it Hipparcos} and ground-based measurements for each proper-motion component,
plotted against the {\it Hipparcos} proper motions $\mu$[HIP] (filled circles:
$\Delta \mu_{\alpha^\ast}$ vs $\mu_{\alpha^\ast}$[HIP], open circles:
$\Delta \mu_\delta$ vs $\mu_\delta$[HIP]). }

\figurenum{6}
\figcaption[]{
(a) Heliocentric radial velocities, $V_{rad}$, for stars
with $\feh \le -0.6$ in the present catalog, as a function of \feh. (b) Total
proper motions, $\mu = (\mu_{\alpha^\ast}^2 + \mu_\delta^2)^{1/2}$, for stars
with $\feh \le -0.6$ as a function of \feh .  Panel (b) clearly indicates the
presence of a substantial number of quite metal-poor stars having small proper
motions, stars which have been absent in previous kinematically selected
samples (see also the discussion in the text).  (c) Distribution of estimated
tangential velocities, $V_t$, as a function of distance from the Sun.  The
solid and dashed lines correspond to lines of constant proper motion of
magnitude 10 mas/yr and 1 mas/yr, respectively. }

\figurenum{7}
\figcaption[]{
Distribution of the complete sample in the $(R,Z)$ plane,
where a Galactocentric distance of $R_\odot=8.5$ kpc has been assumed.
Filled circles and crosses indicate the stars with and without
available proper motions, respectively.}

\figurenum{8}
\figcaption[]{
Local velocity components, $U,V,W$, for the stars in the
revised catalog with complete kinematic information and with abundances $\feh
\le -0.6$, as a function of metallicity.}

\clearpage
%%%%%%%%%%%%%%%%%%%%%%%%%%%%%%%%%%%%
\figurenum{1}
\begin{figure}
\begin{center}
\leavevmode
\epsfxsize=0.9\columnwidth\epsfbox{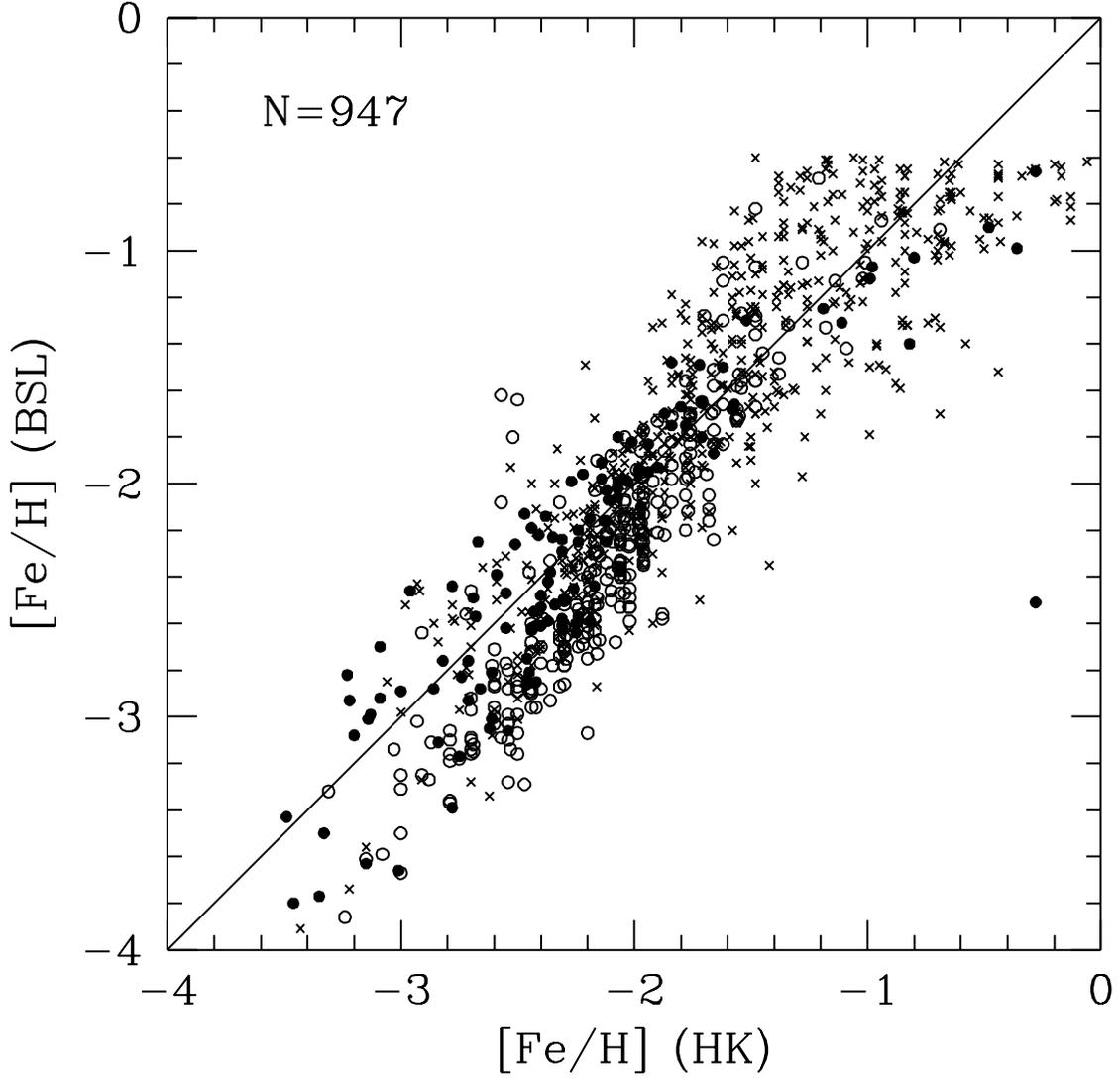}\hfil
\end{center}
\caption{
%\figcaption[fig1.eps]{
Comparison of estimated stellar abundances from the BSL
catalog of Paper I with the revised abundances obtained in the HK survey.  A
one-to-one line is shown. Stars with $0.3 \le (B-V)_o \le 0.5$ are indicated
with open circles, while cooler stars are indicated with filled circles,
respectively.  Stars without available photometry, where abundance estimates
rely on a revised correlation of the Balmer-line index $HP2$ with de-reddened
$(B-V)_o$ color (see Beers et al. 1999) are indicated with crosses.
\label{fig1}}
\end{figure}

\figurenum{2}
\begin{figure}
\begin{center}
\leavevmode
\epsfxsize=0.9\columnwidth\epsfbox{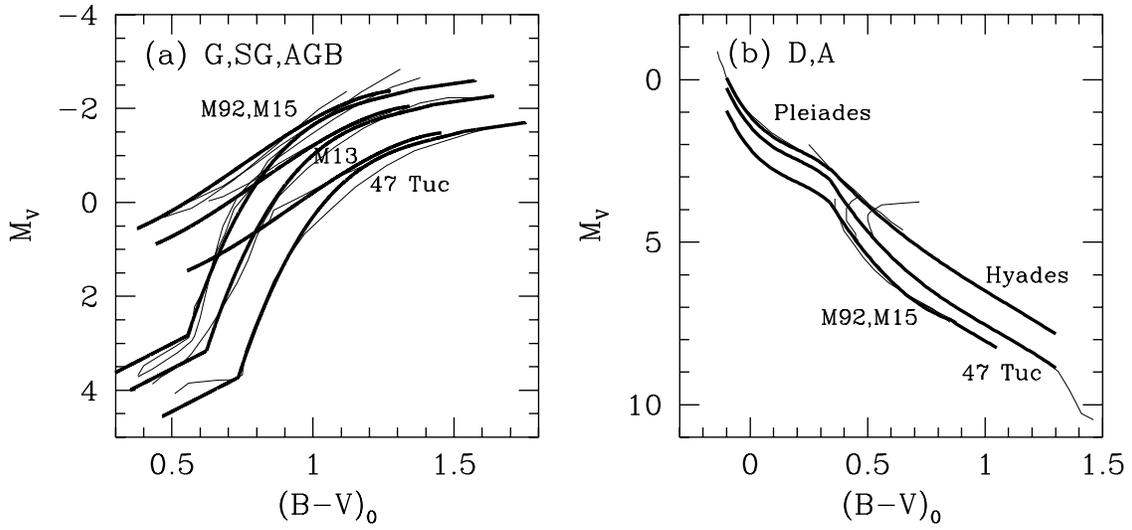}\hfil
\end{center}
\caption{
%\figcaption[fig2.eps]{
(a) $M_V$ vs $(B-V)_o$ relations for G, SG, and AGB stars
in the clusters M92 ($\feh=-2.24$), M15 ($-2.15$), M13 ($-1.65$), and
47~Tuc ($-0.71$) (thin solid lines), together with the adopted $M_V$ vs
$(B-V)_o$ relations at $\feh=-2.20$, $-1.65$, and $-0.71$ (thick solid lines).
(b) The same as in panel (a) but for D stars of M92, M15, 47~Tuc, and Hyades
($\feh=+0.12$).\label{fig2}}
\end{figure}

\figurenum{3}
\begin{figure}
\begin{center}
\leavevmode
\epsfxsize=0.9\columnwidth\epsfbox{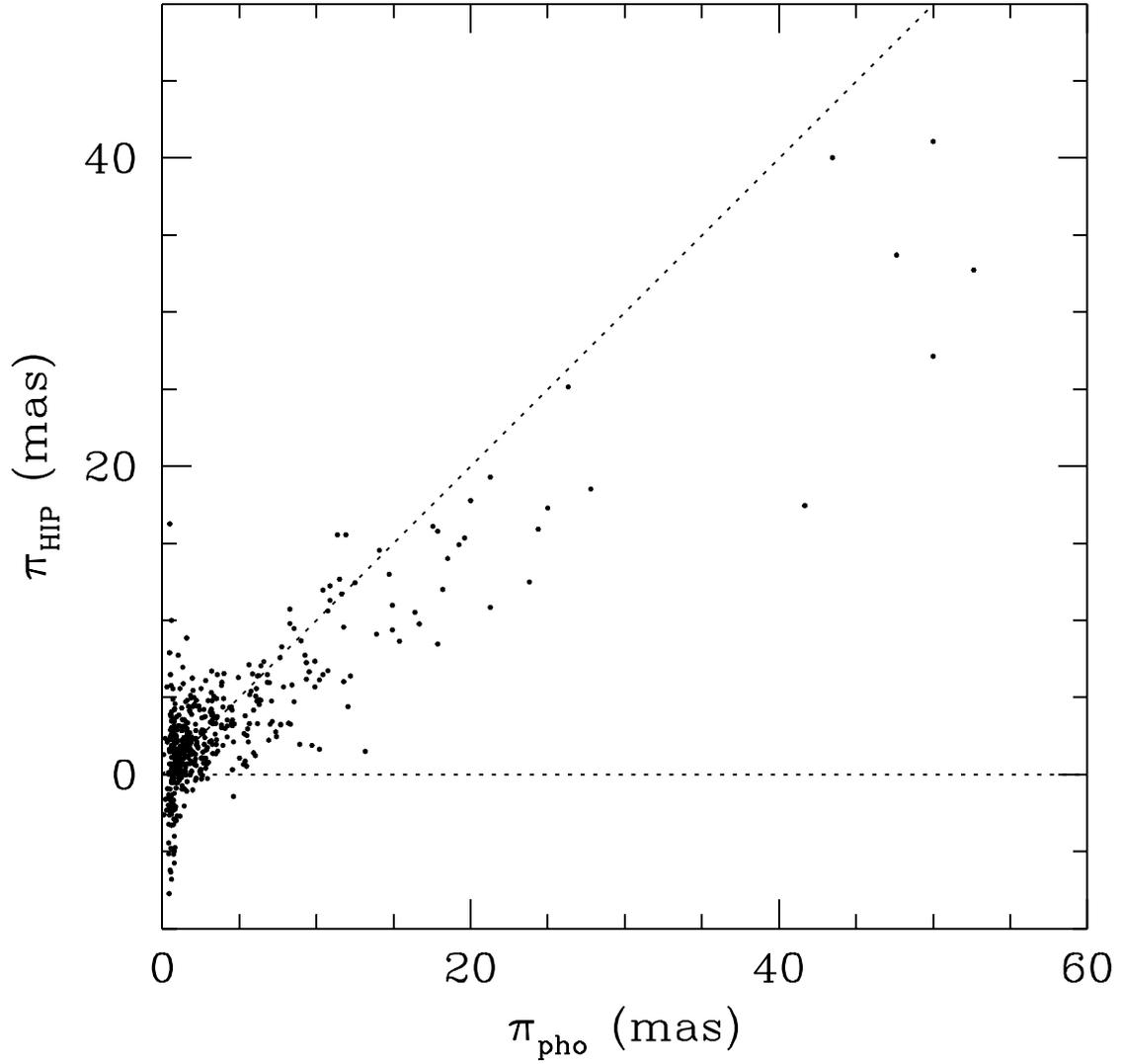}\hfil
\end{center}
\caption{
%\figcaption[fig3.eps]{
{\it Hipparcos} trigonometric parallax $\pi_{HIP}$ versus photometric parallax
$\pi_{pho}$ for 508 {\it Hipparcos} stars having photometric distance estimates
in our catalog.  Units are milli-arc-seconds (mas).  The horizontal dashed 
line indicates $\pi_{HIP} = 0$.  The 45-degree dashed line indicates
$\pi_{HIP} = \pi_{pho}$.
\label{fig3}}
\end{figure}

\figurenum{5}
\begin{figure}
\begin{center}
\leavevmode
\epsfxsize=0.9\columnwidth\epsfbox{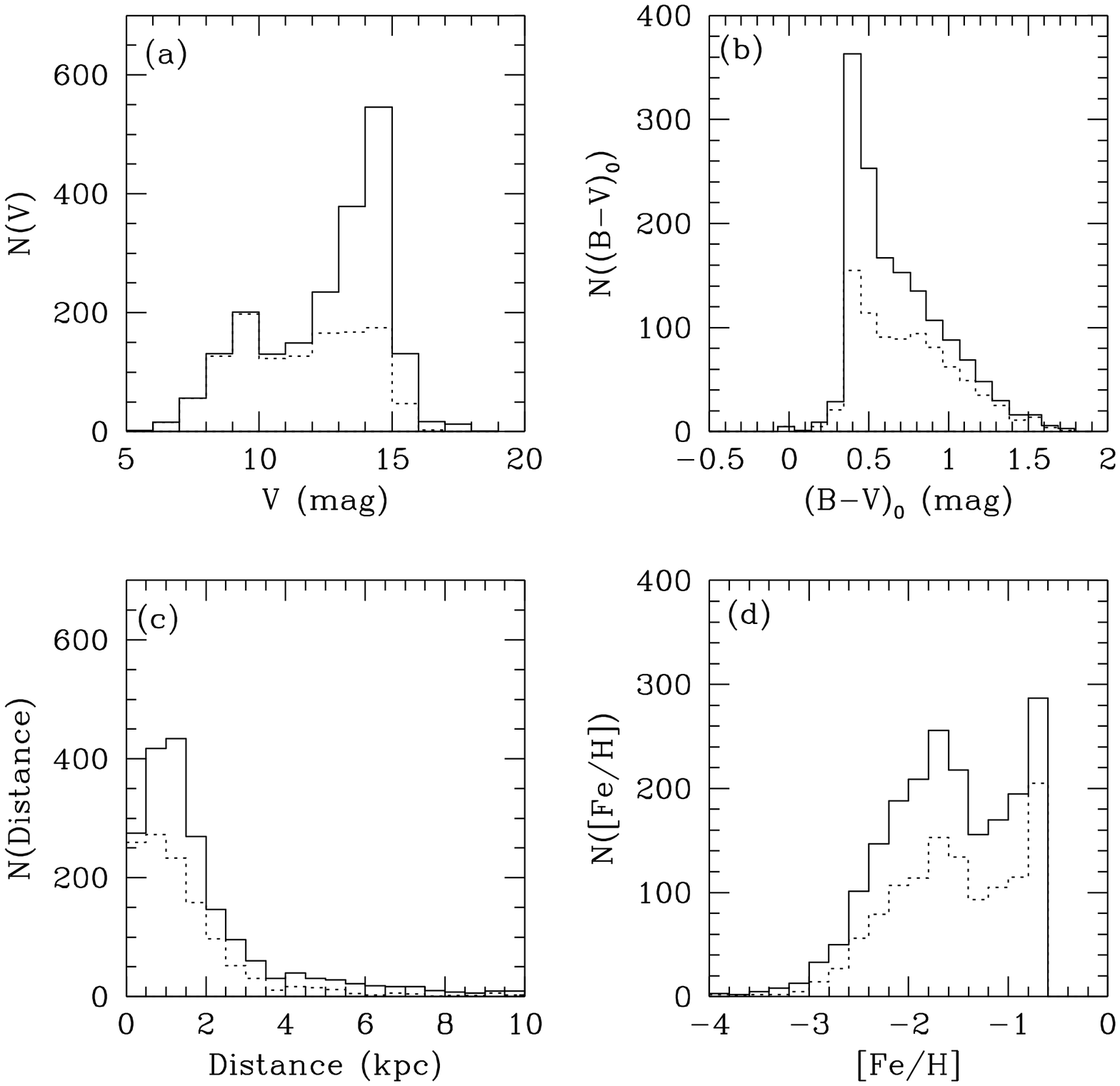}\hfil
\end{center}
\caption{
%\figcaption[fig5.eps]{
Distributions of (a) $V$ magnitudes, (b) $(B-V)_o$
colors, (c) adopted distance estimates, and (d) metal abundances for stars
with $\feh \le -0.6$ in the present catalog. The solid lines apply for the full
sample of stars, while the dashed lines apply for the subsample of stars with
available proper motions.  \label{fig5}}
\end{figure}

\end{document}